# A Superlens Based on Metal-Dielectric Composites

Wenshan Cai, Dentcho A. Genov, and Vladimir M. Shalaev

School of Electrical and Computer Engineering, Purdue University, West Lafayette, IN 47907

Abstract: Pure noble metals are typically considered to be the materials of choice for a near-field superlens that allows subwavelength resolution by recovering both propagating and evanescent waves. However, a superlens based on bulk metal can operate only at a single frequency for a given dielectric host. In this Letter, it is shown that a *composite* metal-dielectric film, with an appropriate metal filling factor, can operate at practically any desired wavelength in the visible and near-infrared ranges. Theoretical analysis and simulations verify the feasibility of the proposed lens.

Negative-index materials (NIMs), which are also referred to as left-handed materials (LHMs), have simultaneously negative real parts of permittivity $\varepsilon$ and permeability $\mu$ so that their refractive index is also negative [1]. These novel materials have recently attracted much attention because of their unprecedented applications. While NIMs do not exist in nature, they have been engineered for microwave wavelengths [2]. Among the most exciting applications discussed so far is the "perfect" lens proposed by Pendry [3], who pointed out that a slab with refractive index $n = -1$ surrounded by air allows the imaging of objects with sub-wavelength precision.

According to Abbe's diffraction limit, conventional lenses based on positive-index materials with curved surfaces are not able to resolve an object's fine details that are smaller than half of the light wavelength $\lambda$. This limitation occurs because the waves with transverse wavenumbers larger than $2\pi n/\lambda$, which carry information about the fine sub-$\lambda$ details of the object, decay exponentially in free space. In a planar NIM slab, however, the evanescent Fourier components can grow exponentially and thus compensate for the exponential decay. Therefore, under ideal conditions, all Fourier components from the object can be recovered at the image plane and a resolution far below the diffraction limit can be achieved [3]. Unfortunately, so far there have been no far-field experimental demonstrations of this super-resolution effect because the conditions for the "perfect lens" are rather severe. In fact, losses (which accompany any resonance-based NIM design) or an impedance mismatch can eliminate the superlensing effect [4,5].

Provided that all of the dimensions of a system are much smaller than the wavelength, the electric and magnetic fields can be regarded as static and independent, and the requirement for superlensing of p-polarized waves (TM mode) is reduced to only $\varepsilon = -\varepsilon_h$, where $\varepsilon_h$ is the permittivity of the host medium interfacing the lens [3]. Although limited to the near-field zone only, this kind of near-field super-lens (NFSL) still allows many interesting applications including biomedical imaging and sub-wavelength lithography. A slab of silver in air illuminated at its surface plasmon resonance (where $\varepsilon = -1$) is a good candidate for such a NFSL. Experiments with silver slabs have already shown rapid growth of evanescent waves [6], submicron imaging [7], and imaging well beyond the diffraction limit [8]. We note, however, that a NFSL can operate only at a single frequency $\omega$ satisfying the lens condition $\varepsilon(\omega) = -\varepsilon_h$, which is indeed a significant drawback of a lens based on bulk metals.

This Letter demonstrates that by employing metal-dielectric composites instead of bulk metals, one can develop a *tunable* NFSL that can operate at any desired visible or near-infrared (NIR) wavelength with the frequency controlled by the metal filling factor of the composite.

The permittivity of metal can be well approximated by the Drude model:

$$\varepsilon_m = \varepsilon'_m + i\varepsilon''_m = \varepsilon_0 - \frac{\omega_p^2}{\omega(\omega + i\Gamma)}, \qquad (1)$$

where $\varepsilon_0$ is the contribution due to interband transitions ($\varepsilon_0 = 5$ for silver*)*, $\omega_p$ is the bulk plasma frequency ($\omega_p = 9$ eV for silver), and $\Gamma$ is the relaxation constant ($\Gamma = 0.02$ eV for silver) [9]. Thus for any given host material with dielectric constant $\varepsilon_h$, the condition $\varepsilon'_m = -\varepsilon_h$ is satisfied only at one particular wavelength; for a silver slab in air, for example, this occurs at $\lambda \approx 340$nm. The operational wavelength $\lambda_{op}$ can be shifted if a host material other than air is used. However, in practice for a particular desired $\lambda_{op}$ it remains a problem to find a host material such that the operating condition $\varepsilon'_m(\lambda_{op}) = -\varepsilon_h(\lambda_{op})$ is exactly fulfilled. In addition, it is difficult to obtain $\lambda_{op}$ beyond the visible range since the value $-\varepsilon'_m(\lambda_{op})$ is too large to match any realistic host medium.

In sharp contrast to pure metal slabs, metal-dielectric composite films are characterized by an effective permittivity $\varepsilon_e$ that depends critically on the permittivities and the filling factors of both the metal and dielectric components. Such composites can be prepared, for example, by evaporation of constituent materials onto a dielectric substrate at ultra-high vacuum [10] with a controllable metal filling factor $p$. As a result, for a given host medium, $\varepsilon_e = \varepsilon_e(\omega,p)$ may have the value of $-\varepsilon_h$ at practically *any* wavelength in the visible and NIR region. The wavelength corresponding to $Re(\varepsilon_e) = -\varepsilon_h$ depends on the structure of the composite and the material constants of the metal and dielectric components in the composite. A schematic for a tunable NFSL is shown in Fig. 1(a).

The optical properties of metal-dielectric composites are well described by the effective medium theory (EMT) [10]. According to the EMT, the effective permittivity $\varepsilon_e$ for a $d$-dimensional composite material comprising metal particles with permittivity $\varepsilon_m$ and a volume filling factor $p$, along with a dielectric component with permittivity $\varepsilon_d$ and a filling factor $1-p$, is given by [11]:

$$p\frac{\varepsilon_m - \varepsilon_e}{\varepsilon_m + (d-1)\varepsilon_e} + (1-p)\frac{\varepsilon_d - \varepsilon_e}{\varepsilon_d + (d-1)\varepsilon_e} = 0. \qquad (2)$$

This is a quadratic equation with the solutions:

$$\varepsilon_e = \varepsilon'_e + i\varepsilon''_e = \frac{1}{2(d-1)}\left\{(dp-1)\varepsilon_m + (d-1-dp)\varepsilon_d \pm \sqrt{[(dp-1)\varepsilon_m + (d-1-dp)\varepsilon_d]^2 + 4(d-1)\varepsilon_m\varepsilon_d}\right\}, \quad (3)$$

where the sign should be chosen such that $\varepsilon''_e > 0$.

The dependence of the effective dielectric permittivity $\varepsilon_e$ on the light wavelength $\lambda$ and on the metal filling factor $p$ is the key to realizing the tunable NFSL. The operational wavelength defined by the condition $Re[\varepsilon_e(p,\lambda_{op})] = -\varepsilon_h(\lambda_{op})$ depends on $p$ and thus can be controlled by varying the metal filling factor. This makes it possible to tune the operating point over a wide wavelength range of interest.

The principle of the tunable NFSL operation is illustrated in Fig. 1(b). The permittivity of silver is given by the Drude model. The effective permittivity of a composite Ag-SiO$_2$ film with a metal filling factor $p$ = 0.85 is calculated by the two-dimensional EMT model. The real part of the effective permittivity $\varepsilon_e$ is smaller than that of pure metal in magnitude; the imaginary part describes a broad, surface-plasmon absorption band resulting from the electromagnetic interactions between individual grains in the composite. The permittivities of air, silicon (Si) and silicon carbide (SiC) are also shown in the figure (we used the tabulated data of [12] and fitted it with functions providing excellent agreement within the visible and NIR range).

For a pure silver slab, the operation wavelengths determined by the condition $Re[\varepsilon_e(p,\lambda_{op})]$ = $-\varepsilon_h(\lambda_{op})$ are indicated in Fig. 1(b) by points A, B, and C for host media of air, SiC and Si, respectively. For the composite NFSL, semiconductor materials like Si and SiC with large $\varepsilon_h$ are beneficial to use as the host materials because they can move $\lambda_{op}$ outside the plasmon absorption band and thus avoid significant losses that are associated with large values of $\varepsilon_e^{''}$ and hence detrimental to the achievable resolution. The imaginary parts of the permittivities of Si and SiC within the wavelength range of interest are negligible and thus do not contribute to losses. As seen in Fig. 1(b), using SiC or Si as the host material, a NFSL with a composite Ag-SiO$_2$ film at $p$ = 0.85 operates at points D and E, respectively, which are both outside the absorption band of the composite.

Thus, for a given host material, one can fabricate a metal-dielectric film with an appropriate filling factor to work for any desired wavelength within a wide wavelength range. For example, with a composite Ag-SiO$_2$ film as the lens and SiC as the surrounding medium, the operational wavelength can be any value to the right of point B (B corresponds to the pure metal with $p$ = 1.0) in Fig. 1(b) until a pre-defined cutoff condition (discussed later) is reached, which determines the long-wavelength and low-$p$ limits for the NFSL operation. Moreover, at the operational point the loss of the lens material can be less than that of pure metal if the resonance peak is avoided. As shown in Fig. 1(b), the curve representing $\varepsilon_e^{''}$ is lower than that of pure metal at the wavelengths corresponding to points D and E. The adverse effect of absorption is less of an issue when semiconductors with high permittivities are used as the host material. This is because such materials provide a better spatial resolution, which is approximately proportional to $1/ln(|\varepsilon^{'}/\varepsilon^{''}|)$ for a lens material with permittivity $\varepsilon = \varepsilon^{'}+i\varepsilon^{''}$ [13].

Fig. 2(a) shows the required metal filling factor $p$ for superlens operation using an Ag-SiO$_2$ composite lens. The filling factors were found from the superlens equation $Re[\varepsilon_e(p,\lambda_{op})] = -\varepsilon_h(\lambda_{op})$ for different wavelengths with Si or SiC as the host medium. For each kind of host material, the lower limit of the operational wavelength range corresponds to the pure metal ($p = 1$) case. A lower metal filling factor is required for a longer operational wavelength. For very long wavelengths, the required filling factor for the composite approaches the percolation threshold where the broad resonance peak in $\varepsilon_e^{''}(\lambda)$ reduces the super resolution effect. In Fig. 2 the upper limit of the possible wavelength range is determined so that $Im(\varepsilon_e)/Re(\varepsilon_h) = 0.1$ at the longer wavelength end of the operational range. Note that the criterion we use here is a very conservative one. For a silver lens working in air as first proposed in [3] or polymethyl methacrylate (PMMA) as studied in [7,8], the value of $Im(\varepsilon_e)/Re(\varepsilon_h)$ was as large as 0.4 while sub-wavelength resolution was still achievable. As seen in Fig. 2(a), with an Ag-SiO$_2$ composite as the lens material the operation ranges are 0.47 – 0.67 μm for SiC host material and 0.61 – 1.10 μm for Si host material. Therefore, combining the results of the two host media we can achieve a possible operational wavelength range of 0.47 – 1.10 μm, which covers nearly the whole visible spectrum and the shorter part of NIR band. Note that the performance of the NFSL will not be spoiled even the photon energy at the operational wavelength is larger than the band gap of the host semiconductor because the total thickness of the host medium is only a few tens of nanometers. The operational range can be expanded even further by using other host media or other constituent materials for the metal-dielectric composite. The value of $Im(\varepsilon_e)/Re(\varepsilon_h)$ as a function of the operational wavelength is plotted in Fig. 2(b).

To illustrate the imaging ability of the proposed tunable NFSL based on metal-dielectric films, we calculate the optical transfer function (OTF) of the system and the image formed at the imaging plane for a given object. For simplicity we assume invariance along the y direction for the whole system (coordinates are shown in Fig. 1(a)). The OTF from the object plane to the image plane is given by

$$OTF(k_x) = T_p(k_x) \exp(ik_z^{(1)} d_1) \exp(ik_z^{(2)} d_2), \qquad (4)$$

where $T_p(k_x)$ is the transmission coefficient of the slab for the p-wave (TM mode). The index "1" represents the material between the object plane and the lens and the index "2" represents the material between the lens and the image plane. The transmission coefficient is given by [14]:

$$T_p(k_x) = \frac{4(k_z^{(1)}/\varepsilon_1)(k_z^{(e)}/\varepsilon_e)\exp(ik_z^{(e)}d)}{(k_z^{(1)}/\varepsilon_1 + k_z^{(e)}/\varepsilon_e)(k_z^{(2)}/\varepsilon_2 + k_z^{(e)}/\varepsilon_e) - (k_z^{(1)}/\varepsilon_1 - k_z^{(e)}/\varepsilon_e)(k_z^{(2)}/\varepsilon_2 - k_z^{(e)}/\varepsilon_e)\exp(2ik_z^{(e)}d)}, \qquad (5)$$

where $k_z^{(i)} = \sqrt{\varepsilon_i k_0^2 - k_x^2 - k_y^2}$.

As an example, the performance of the Ag-SiO$_2$ NFSL with SiC as the host and operating at 632.8 nm is illustrated in Fig. 3. At this wavelength the host permittivity $\varepsilon_h = 6.94$, the effective permittivity of the lens $\varepsilon_e = -6.94+i0.31$, and the required metal filling factor given by the relation $Re[\varepsilon_e(p,\lambda_{op})] = -\varepsilon_h(\lambda_{op})$ is $p = 0.82$. The thickness of the lens is chosen to be $d = 20$ nm. The modulation transfer function (MTF), which is defined as $MTF(k_x) = |OTF(k_x)|^2$, represents a useful way to evaluate the imaging ability of a NFSL [15]. In Fig. 3(a) the MTF of the imaging system (solid line) together with the MTF from the object plane to the image plane without the lens are plotted as functions of the transverse wavevector $k_x$. For the perfect lens ($n = -1$; metal slab in vacuum), $MTF(k_x) = 1$ for all $k_x$ and thus a perfect image can be produced. For free space, as shown by the dashed line in Fig. 3(a), $MTF(k_x) = 1$ for propagating waves (when $k_x/k_0 < 1$) and it decays exponentially for evanescent waves (when $k_x/k_0 > 1$). For the composite NFSL with the parameters given above, the MTF can maintain a value comparable to unity for a range of $k_x$ up to $15k_0$, which indicates a resolution of about $\pi/15k_0 = \lambda/30$ can be obtained. If the Fourier transform of the field distribution in the object plane is $E_{obj}(k_x)$, then at the image plane each Fourier component is $E_{image}(k_x) = OTF(k_x)E_{obj}(k_x)$ and the inverse Fourier transform produces the field distribution in the image plane. Fig. 3(b) illustrates the simulated result of the image of a pair of slits of width $d$ and center-to-center separation $2d$. The result shows that the composite lens is capable of reconstructing the object at the image plane. Without the lens, the pair of slits cannot be resolved, as shown by the dashed curve in Fig. 3 (b). Note that the procedure to calculate the MTF and the image as described above is exact (no quasi-static or other approximations were used). The only assumption is y-invariance (such as in a slit pair) and p-wave illumination.

The use of the effective parameters for a composite film implies that the object-film and film-image distances exceed the typical sizes of nanoparticles. When a large distance between the object and the image plane is required, one can use a multi-layer structure with interleaved negative (metal-dielectric composite) and positive (host) slices, which can be regarded as a series of cascaded lenses. The detrimental effects such as loss are reduced in the multi-layer structure [16]. Moreover, the interaction between metal clusters in neighbor metal-dielectric films may result in optical magnetism, enabling a far-zone superlens that can image objects that are well separated from the lens (to be published elsewhere).

For all of the results up to this point, the effective permittivity of the metal-dielectric composite is calculated using the two-dimensional EMT, which provides a simple analytical way to evaluate the effective properties of the composite film. To assure that the composite NFSL does give acceptable resolution, we re-visit the properties of the composite film using the block-elimination (BE) method. The BE method is an exact numerical approach to calculate the effective parameters and local-field distribution of a metal-dielectric film. The detailed procedure is demonstrated in [17]. Compared to the results by EMT, the resonance range of the effective permittivity calculated by the BE method has a longer tail. Therefore, the value of $\varepsilon_e''$ at the operational wavelength of the NFSL obtained by the BE method is somewhat larger compared to the EMT results. The imaginary part of permittivity of an Ag-SiO$_2$ composite with different metal filling factors calculated by the EMT and the BE method are illustrated in Fig. 4(a) and 4(b), respectively; the two calculations are in reasonable agreement. Because the BE algorithm is time-consuming, it is not feasible to calculate the exact $p = p(\lambda)$ relationship of the composite NFSL using this method. However, using the $p = p(\lambda)$ relation given in Fig. 2(a) as a starting point, one can verify that the $\varepsilon_e''$ obtained by the exact BE method is indeed acceptable for the composite NFSL to provide the superlensing effect.

There is no unambiguous way to define the resolution of an imaging system. Here we use a modified version of Rayleigh's criterion to evaluate the limit of resolution for the proposed composite NFSL system. Similar to the example in Fig. 3(b), we consider a NFSL system where the lens thickness is $d$ and the object is a pair of slits of width $d$ and center-to-center separation $2d$. At the image plane, the two slits are regarded as barely resolvable if the intensity at the midpoint of the slit pair is $8I_0/\pi^2 = 0.811I_0$, where $I_0$ is the maximum intensity. This criterion provides an upper bound for the detrimental effects on NFSL performance such as loss, impedance mismatch, retardance, and others. Considering an Ag-SiO$_2$ composite NFSL of $d = 20$nm with Si or SiC as the surrounding medium, the upper bound of $\varepsilon_e''$ determined by the Rayleigh criterion is shown in Fig. 4(c), together with the values of $\varepsilon_e''$ within the operational wavelength range obtained by the BE method using linear interpolation. We see that the $\varepsilon_e''$ calculated by the exact two-dimensional BE method is still far below the upper bound of $\varepsilon_e''$ defined by the Rayleigh criterion, which verifies that the composite NFSL can give near-field super-resolution.

In summary, we have proposed a tunable near-field superlens based on a metal-dielectric composite. Our analysis and simulations show that with a high-permittivity host material, the composite lens can be engineered with a suitable metal filling factor to operate at practically any desired wavelength in the visible and near-infrared range. The tunable NFSL may find applications in sub-wavelength imaging, nanolithography, non-contact bio-molecule sensing and spectroscopy.

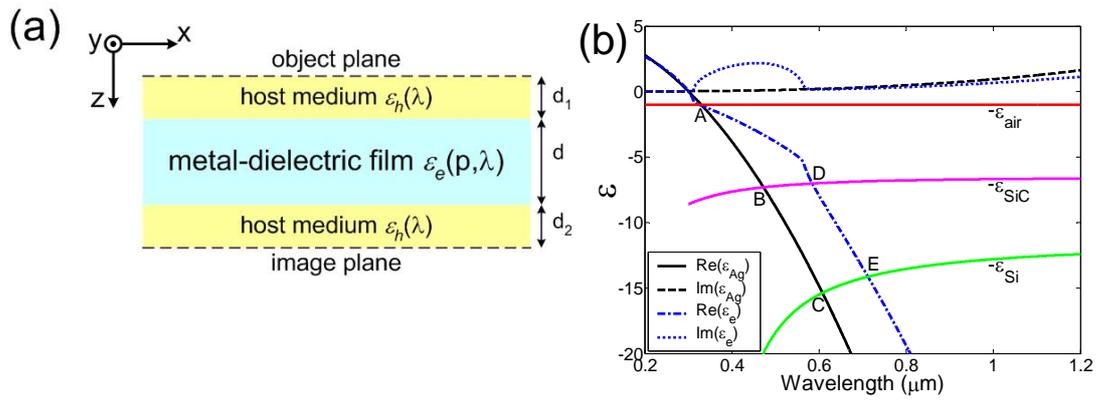

Fig. 1 (color online). (a) Schematic of the tunable NFSL based on a metal-dielectric composite. (b) Principle of NFSL operation. The composite used is an Ag-SiO$_2$ film with metal filling factor $p$ = 0.85. The operational points are: A – silver lens with air host ($\lambda_{op}$ = 340nm); B – silver lens with SiC host; C – silver lens with Si host; D – composite lens with SiC host; E – composite lens with Si host.

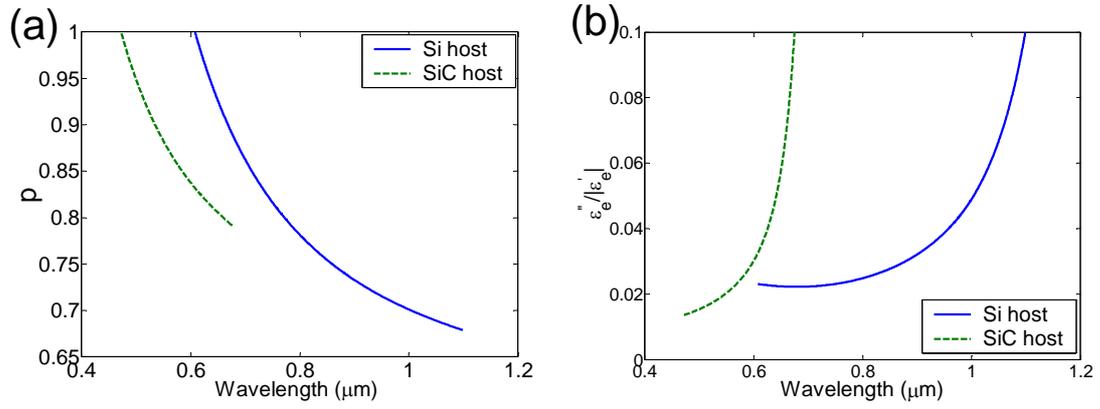

Fig. 2 (color online). Performance of an Ag-SiO$_2$ composite lens with Si or SiC as the host medium. (a) The required metal filling factor $p$ for different wavelengths. (b) The value of $Im(\varepsilon_e)/Re(\varepsilon_h)$ for different wavelengths.

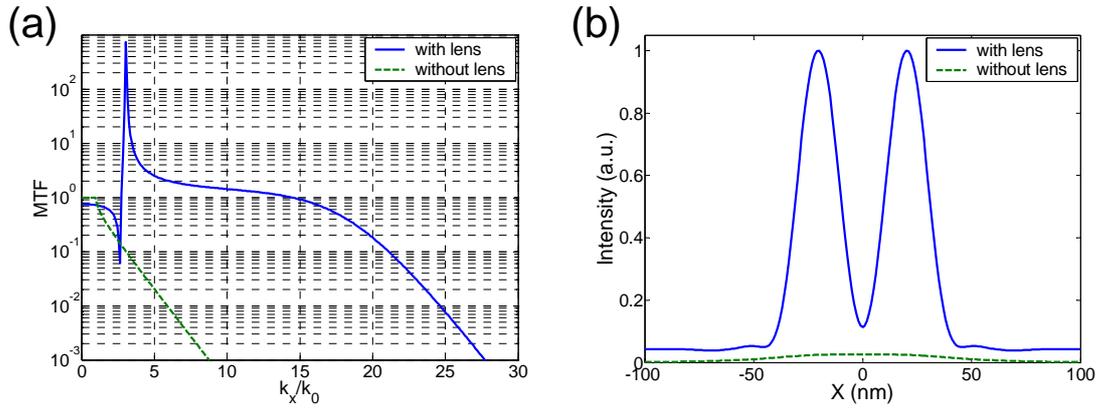

Fig. 3 (color online). The imaging ability of a 20 nm Ag-SiO$_2$ composite lens with SiC as the host medium working at 632.8 nm. (a) The MTF of the imaging system as a function of the transverse wavevector $k_x$. (b) The simulated result of the image of a pair of slits of width $d$ and center-to-center separation $2d$.

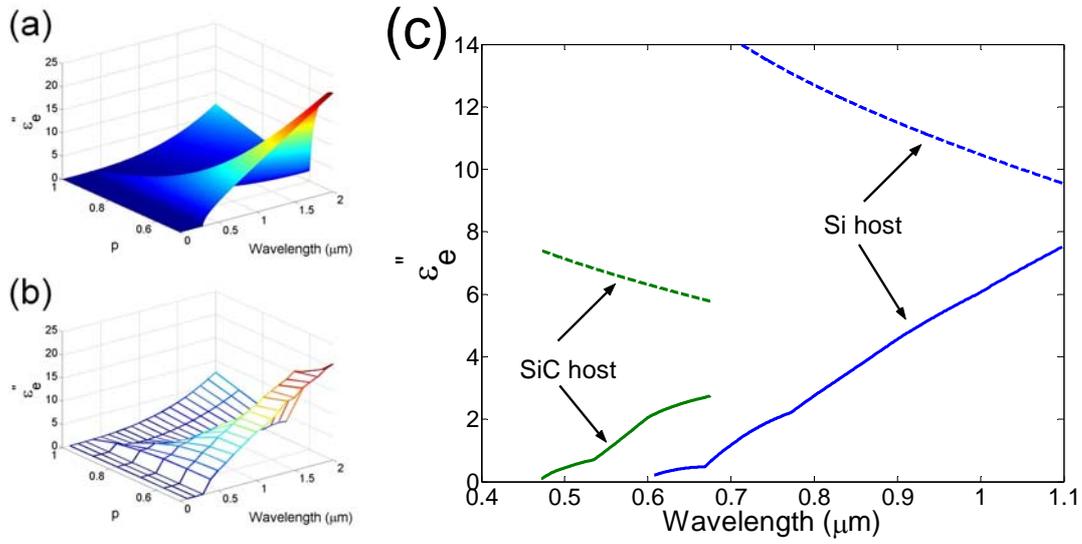

Fig. 4 (color online). (a) The imaginary parts of the effective permittivity $\varepsilon_e''$ of an Ag-SiO$_2$ composite calculated by the EMT. (b) $\varepsilon_e''$ calculated by the BE method. (c) The upper bound of $\varepsilon_e''$ determined by Rayleigh criterion (dashed) and the $\varepsilon_e''$ at the operational wavelength calculated by the BE method (solid). The NFSL system is a 20 nm Ag-SiO$_2$ composite with Si or SiC as the host medium. The object is the same as that in Fig. 3(b).